\documentclass{llncs}

\usepackage{amssymb}
\usepackage{graphicx}
\usepackage{multicol}
\usepackage{url}

\begin{document}
\title{On weak universality of three-dimensional Larger than Life cellular automaton}
\titlerunning{A universal 3D LtL}  
%
\author{Katsunobu Imai \and Kyosuke Oroji \and Tomohiro Kubota}
\authorrunning{Imai et al.} 
%
\tocauthor{Katsunobu Imai, Kyosuke Oroji and Tomohiro Kubota}
\institute{Hiroshima University, Hiroshima, Japan\\
\email{imai@hiroshima-u.ac.jp}}

\maketitle              

\begin{abstract}
Larger than Life cellular automaton (LtL) is a class of cellular automata and is a generalization of the game of Life by extending its neighborhood radius. We have studied the three-dimensional extension of LtL. In this paper, we show a radius-$4$ three-dimensional LtL rule is a candidate for weakly universal one. 

\end{abstract}

\section{Introduction}

A class of two-dimensional cellular automata, {\em Larger than Life} (LtL)~\cite{Evans2001}, is a natural generalization of the game of Life~\cite{BCG1982} by extending the radius of its neighborhood. 
A lot of interesting patters such as {\em bugs} are found in LtL~\cite{Evans2003}. 
A bug is a kind of {\em glider} or {\em spaceship}  patterns found in the game of Life~\cite{Gardner1970,BCG1982}. 
An radius-$5$ LtL rule, {\em Bosco}~\cite{Evans2001} has high computing efficiency employing the function of bugs and the other coherent structures~\cite{Evans2004}. 

As far as the game of Life, there are several studies about three-dimensional extensions~\cite{Bays1987,Bays1991}. 
Although there are many combination of the ranges of birth and survival, Bays found some better candidates as the three-dimensional ``Life"  and found several glider patterns~\cite{Bays1987,Bays1994}. Although the computational universality of (two-dimensional) Life was shown more than 35 years ago~\cite{BCG1982}, there is no related study of universality of three-dimensional Life, to the best of our knowledge.

We have studied LtL rules in three-dimension.  
We showed a methodology to design period-$1$ bugs and constructed several period-$1$ bugs in the cases of radii from $3$ to $7$ and we also found several stable patterns and blinker patterns~\cite{IMIM2010}. 
At the time, we also tried to find some functional patterns such as turning a bug or copying a bug by colliding a bug to blinkers or stable blocks. Because they are thought to be inevitable for embedding a computing function to a cellular space. But the trial was not fruitful. Because there was two reasons: (1) the number of found bugs was too small, (2) period-$1$ bugs are too `solid' to recover the shapes after some reactions cased by the collisions of another bug, stable block or blinker. 

In the process of constructing these bugs and the simulation of collisions of them, we also found a period-$10$ bug in the case of radius $4$~\cite{IMIM2010}. 
We noticed that the bug has a kind of stabilizing property to recover the original shape of the bug even starting from a pattern which is not any pattern of the bug. 
Unfortunately the bug did not have any useful function, we have been trying to find bugs with bit longer periods. 

We have finally found a radius-$4$ three-dimensional LtL rule as a weakly universal candidate, i.e, starting from a periodic infinite initial configuration, it is possible to simulate a universal machine (the rule 110 cellular automaton can be simulated~\cite{Cook2004}). 

\section{Larger than Life cellular automata}

Larger than Life (LtL)~\cite{Evans2001} $L$ is 
a natural generalization of the game of Life and defined by $L=(r,\beta_1,\beta_2,\delta_1,\delta_2)$, where $r$ is the radius of its 
neighborhood, $[\beta_1,\beta_2]$ is the range of the total number of live 
neighborhood cells for giving birth to the central focus cell ($0 \rightarrow 1$), $[\delta_1,\delta_2]$ 
is the range for survival of the central focus cell ($1 \rightarrow 1$).
The shape of its neighborhood is the square of the length $2r+1$, thus the neighborhood 
includes $(2r+1)^2$ cells. 
Note that the game of Life can be denoted by $(1,3,3,3,4)$. 

In the case of LtL, Evans~\cite{Evans2001} found a lot of special glider like patterns called a {\em bug}. 
A bug is a pattern which crawls the cellular space and the same pattern repeatedly appears at a certain {\em displaced} position $d=(d_1,d_2)$ after a certain {\em period} $\tau$. 

\begin{figure}[h]
\begin{center}
\includegraphics[scale=0.3]{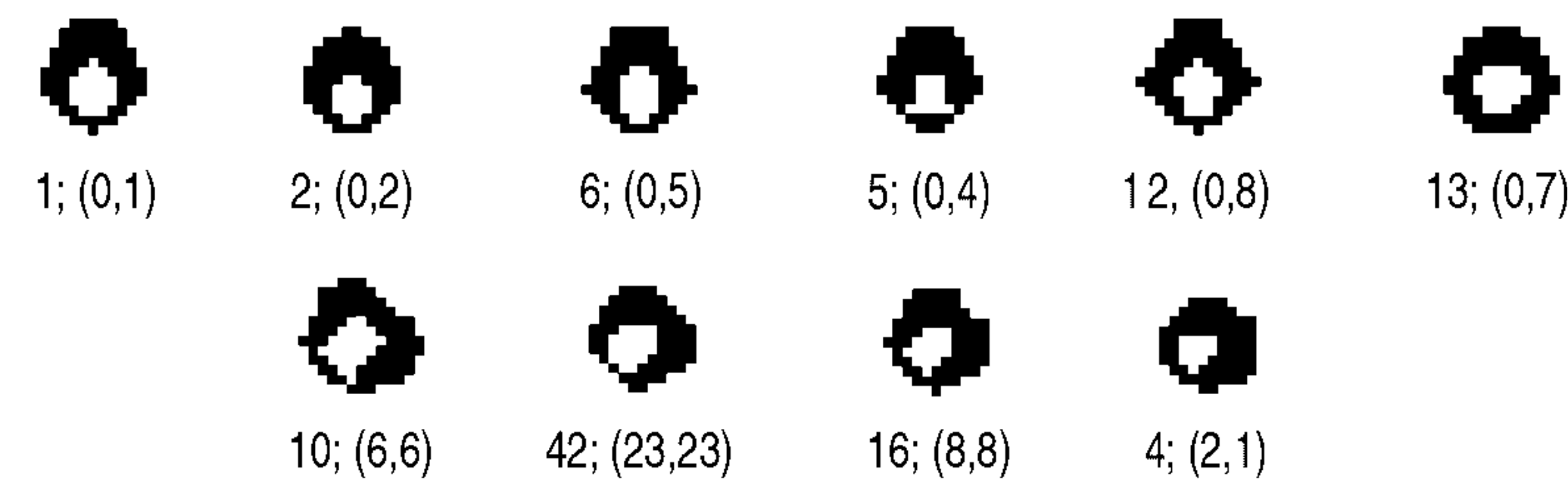}
\caption{Family of bugs $\tau;(d_1,d_2)$ supported by Bosco's rule $(5,34,45,34,58)$~\cite{Evans2001}.}
\label{fig:bugs-evans}
\end{center}
\end{figure}

LtL can be also defined in three-dimension in the same way, except the number of neighborhood cells is $(2r+1)^3$. It is extremely large, for example, $729$ in the case of radius-$4$. A displacement vector of a bug needs to be also modified to three-dimension. 

\section{Three-dimensional LtL rule $(4,102,133,102,142)$}
\subsection{Bug, blinkers and basic modules}

In this section we show that the three-dimensional LtL rule, $L=(4,102,133,102,142)$, has interesting patterns such as a bug, blinkers, and basic modules which are useful for executing computations by embedding a logical circuit. 

\begin{figure}[htbp]
\begin{center}
\includegraphics[width=0.6\linewidth]{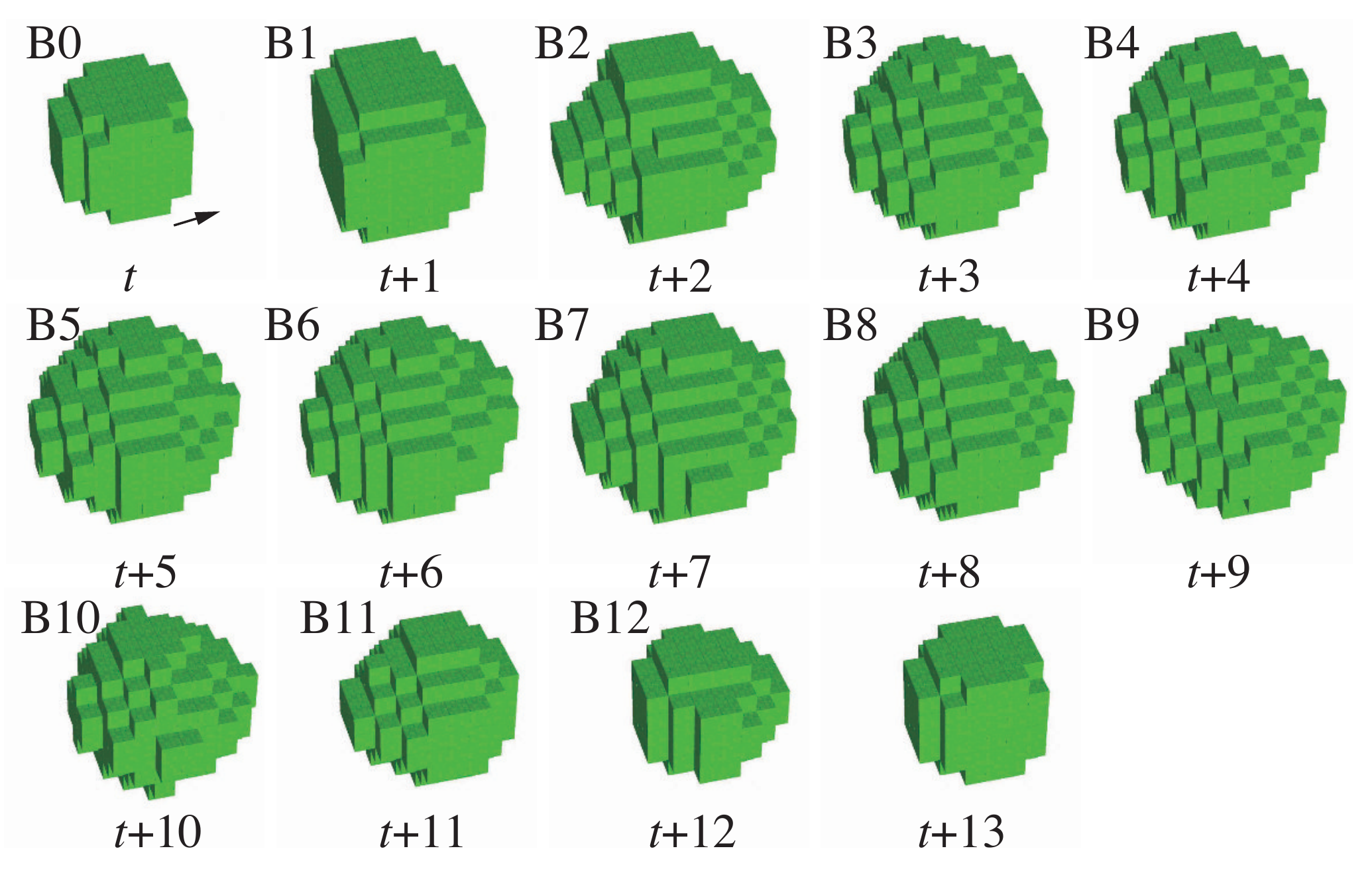}
\caption{The evolution of a bug B in a period.}
\label{fig:bug-B-patterns}
\end{center}
\end{figure}

\begin{figure}[htbp]
\begin{center}
\includegraphics[width=0.5\linewidth]{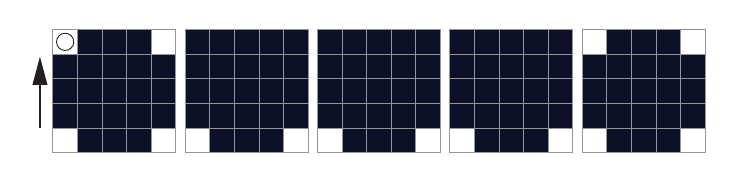}
\caption{The cross-sectional view of a bug B0.}
\label{fig:B0-schematic}
\end{center}
\end{figure}

There is a period-$13$ bug B which advances two cells in 13 time steps, i.e., $\tau_B=13, d_B=(0,2,0)$. Fig.~\ref{fig:bug-B-patterns} shows the evolution of the patterns in a period. 
Fig.~\ref{fig:B0-schematic} shows the detailed cross-sectional pattern of B0 of the bug B. In the following, the coordinate of the position of the black circle is used to identify the position of B0.

\begin{figure}[htbp]
\begin{center}
\includegraphics[width=0.6\linewidth]{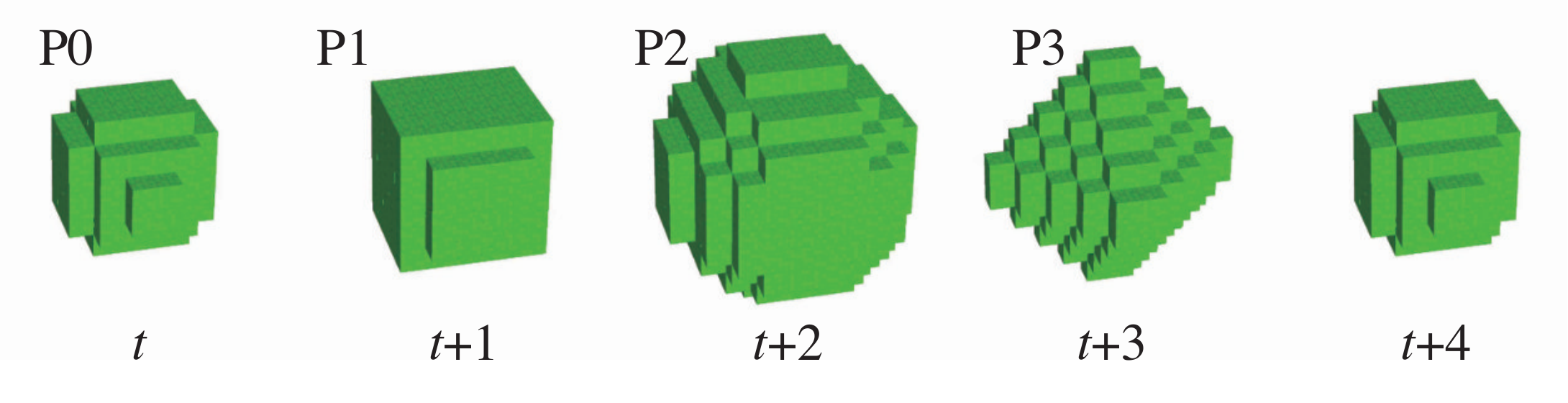}
\caption{The evolution of period-$4$ brinker P in a period.}
\label{fig:bilinker-P-patterns}
\end{center}
\end{figure}

We also found two blinkers P and Q.
P is a period-4 blinker. Fig.~\ref{fig:bilinker-P-patterns} shows the evolution of the patterns in a period. 
\begin{figure}[htbp]
\begin{center}
\includegraphics[width=0.7\linewidth]{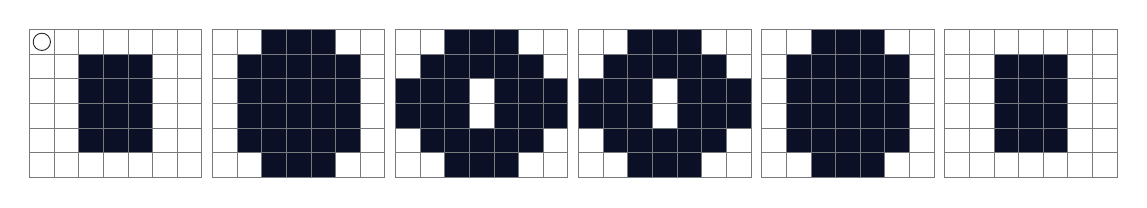}
\caption{The cross-sectional view of P2.}
\label{fig:P-schematic}
\end{center}
\end{figure}
Fig.~\ref{fig:P-schematic} shows the cross-sectional pattern P2 of the blinker P in Fig.~\ref{fig:bilinker-P-patterns}.

\begin{figure}[htbp]
\begin{center}
\includegraphics[width=0.6\linewidth]{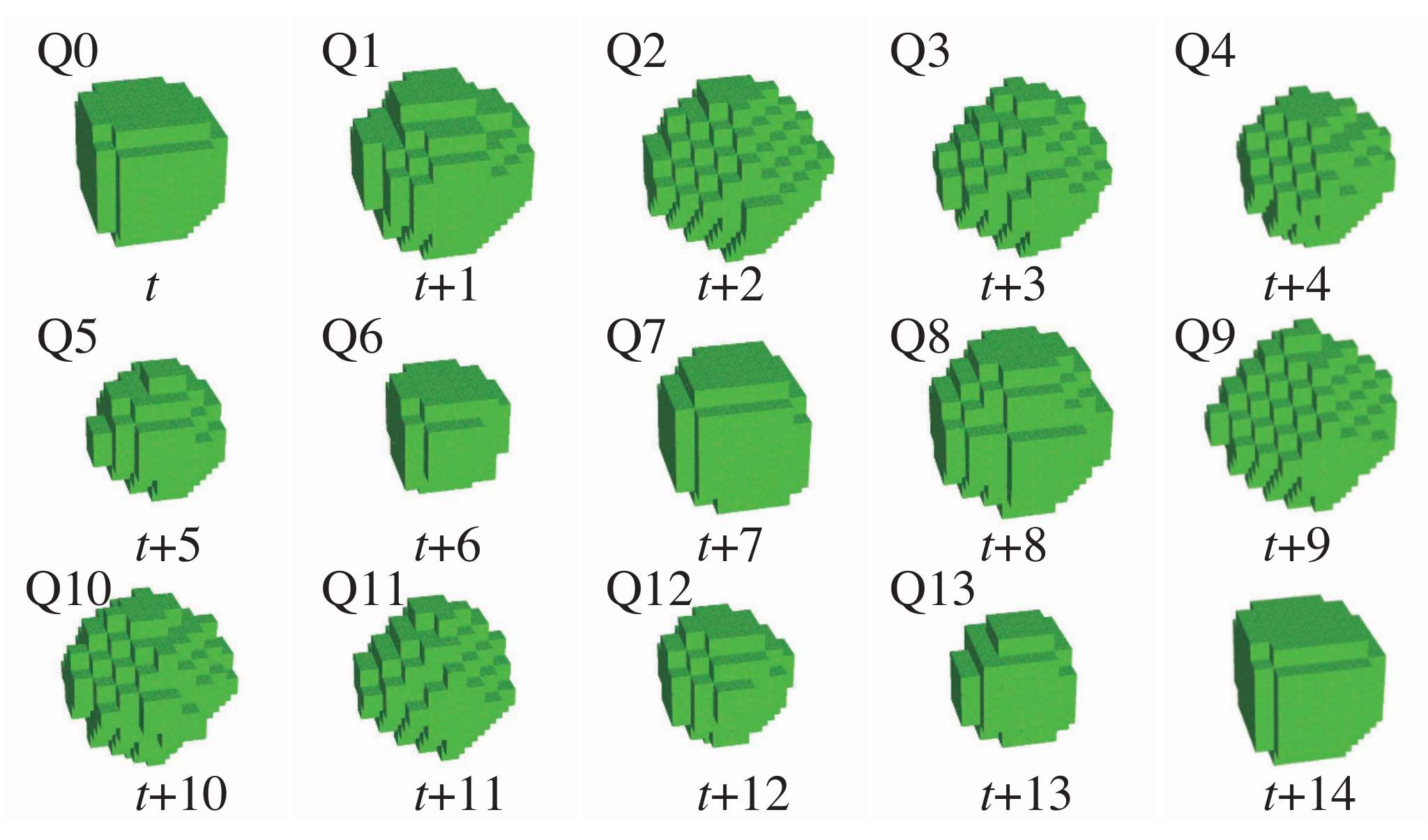}
\caption{The evolution of period-$14$ brinker Q in a period.}
\label{fig:bilinker-Q-patterns}
\end{center}
\end{figure}

\begin{figure}[h]
\begin{center}
\includegraphics[width=0.9\linewidth]{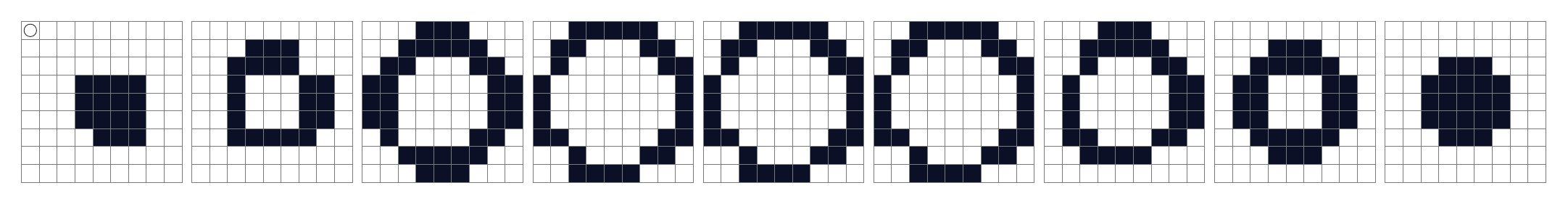}
\caption{The cross-sectional view of Q1.}
\label{fig:Q-schematic}
\end{center}
\end{figure}

Q is a period-14 blinker. Fig.~\ref{fig:bilinker-Q-patterns} shows the evolution of the patterns in a period. In contrast to the blinker P, the blinker Q is asymmetric in the structure. 
Q$i$ and Q$i+7 \ (0<i<6)$ are symmetric and both sequences are alternatingly appeared. 
Fig.~\ref{fig:Q-schematic} shows the cross-sectional pattern of Q1 of the blinker Q in Fig.~\ref{fig:bilinker-Q-patterns}. 

\begin{figure}[h]
\begin{center}
\includegraphics[width=0.6\linewidth]{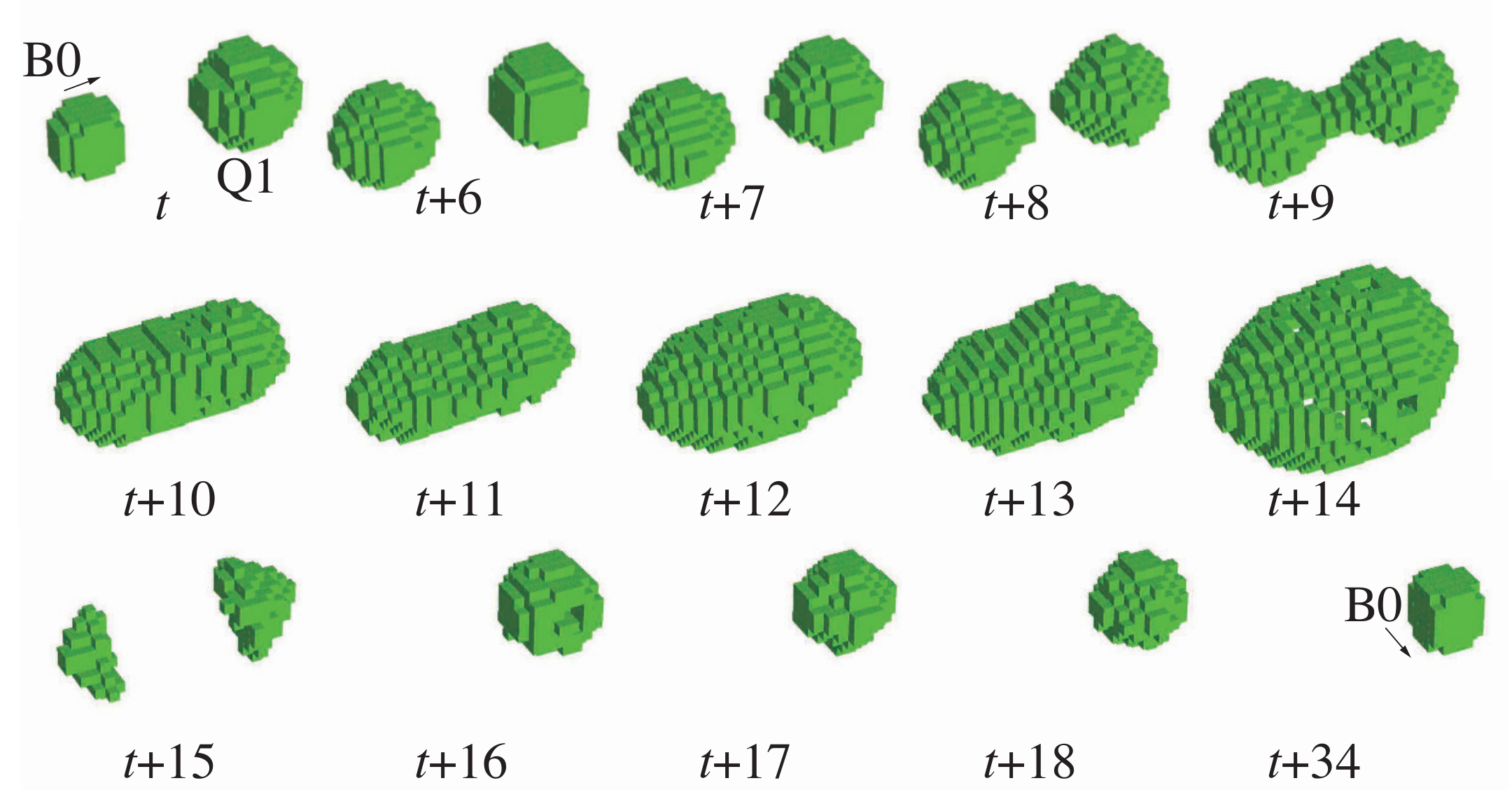}
\caption{A process of turning a bug.}
\label{fig:bug-turn}
\end{center}
\end{figure}

\begin{figure}[h]
\begin{center}
\includegraphics[width=0.4\linewidth]{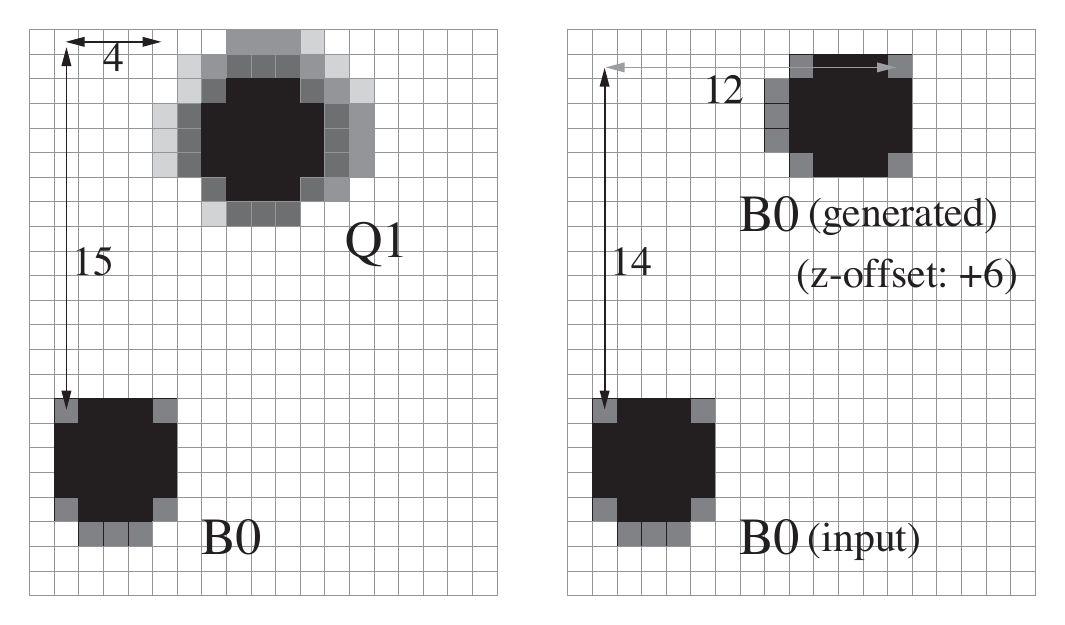}
\caption{A layout of turning configuration in $xy$-plane and the relative positions of input and output bugs.}
\label{fig:turn-schematic}
\end{center}
\end{figure}

A collision of a bug B to a blinker Q, with a special offset and phase of a period, realizes a {\em turning module} of the bug B. 
Fig.~\ref{fig:bug-turn} shows a process of the right turning. An input bug B0 changes the direction. 
Place B0 at $(0,0,0)$ and Q1 at $(4,14,6)$ as an initial configuration, after $34$ steps, a right-turned B0 is generated at $14,12,6$ as illustrated in Fig.~\ref{fig:turn-schematic}. 
At time $17$ in Fig.~\ref{fig:bug-turn}, the shape of output pattern is almost recovered but it takes several more steps to converge the exact shape of the bug B. 

\begin{figure}[h]
\begin{center}
\includegraphics[width=0.6\linewidth]{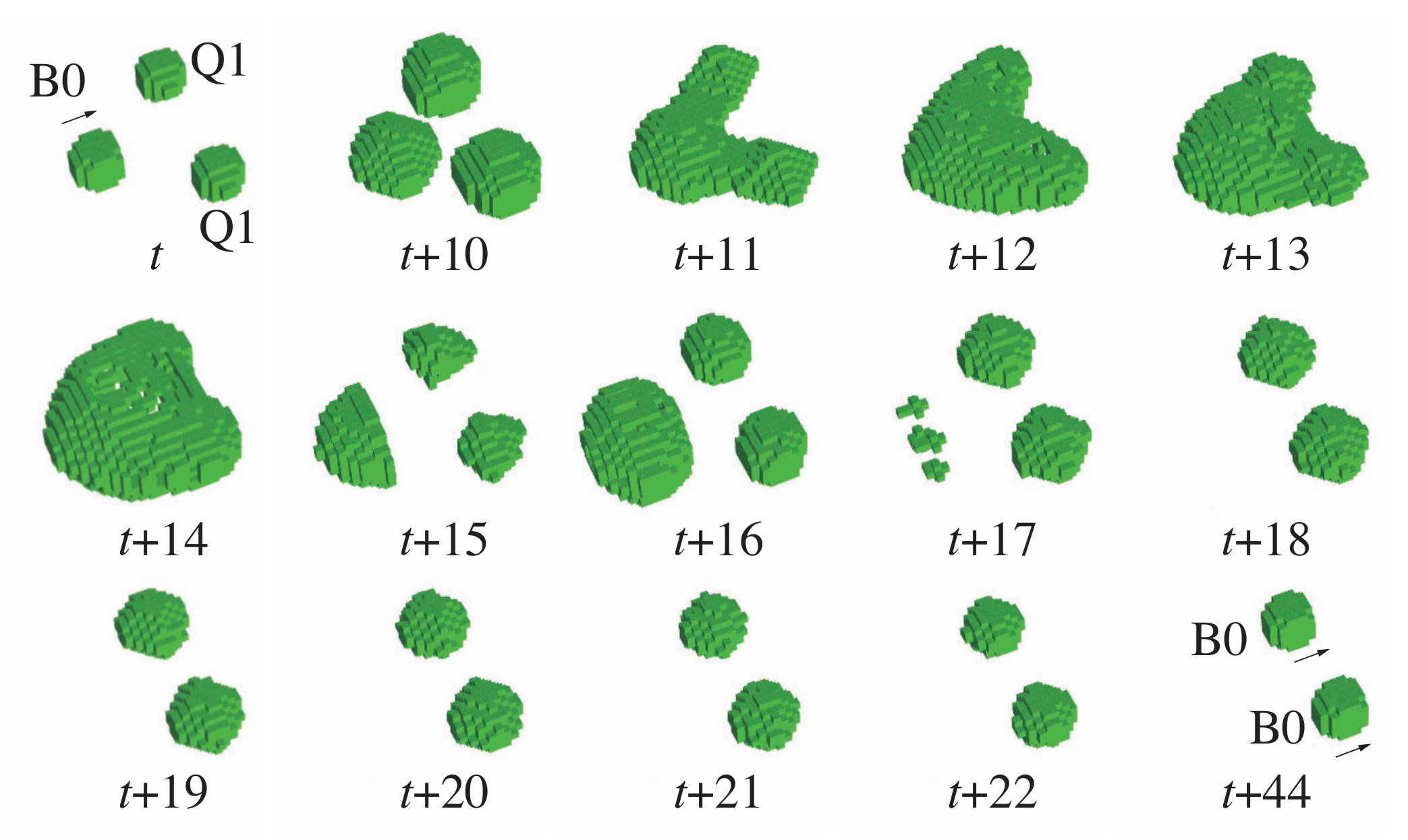}
\caption{A process of duplicating a bug.}
\label{fig:bug-copy}
\end{center}
\end{figure}

\begin{figure}[h]
\begin{center}
\includegraphics[width=0.5\linewidth]{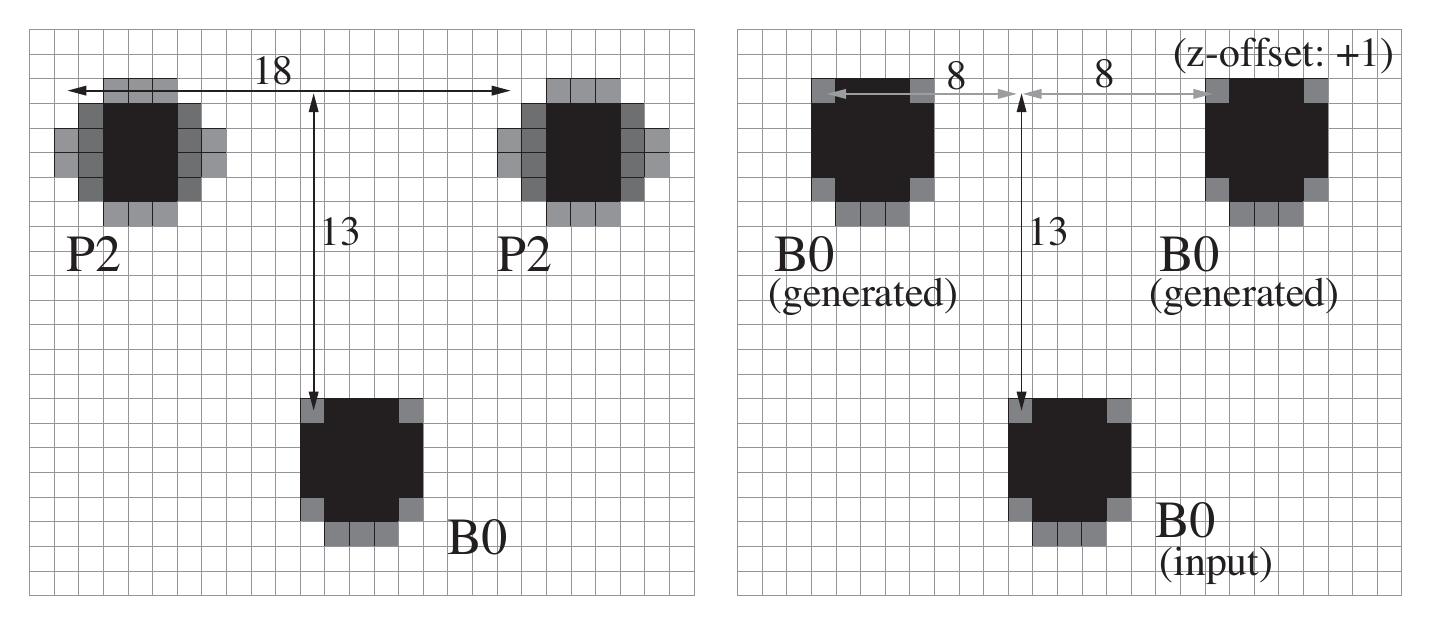}
\caption{A layout of duplicate configuration in $xy$-plane and the relative positions of input and output bugs.}
\label{fig:dupbug-schematic}
\end{center}
\end{figure}

A {\em duplicator} of a bug B can be realized by a collision of B to two blinkers, Q, with an offset and a phase. 
Fig.~\ref{fig:bug-copy} shows a duplicating process of a bug B. A bug B0 as an input, two B0 patterns are generated as outputs. 
As in Fig.\ref{fig:dupbug-schematic}, $B0$ at $(0,0,0)$ and two Q1 are placed at $(\pm 9, 13, 1)$ as an initial configuration, two $B0$ patterns finally appear at the  positions $(\pm 8,13,1)$ after $44$ time steps.

Note that these collisions are destructive and each function is only used once. 

\subsection{Logical gates}

We shows that a NOT gate module and an AND gate module can be realized by combining a duplicator and a turning module. 

\begin{figure}[h]
\begin{center}
\includegraphics[width=0.7\linewidth]{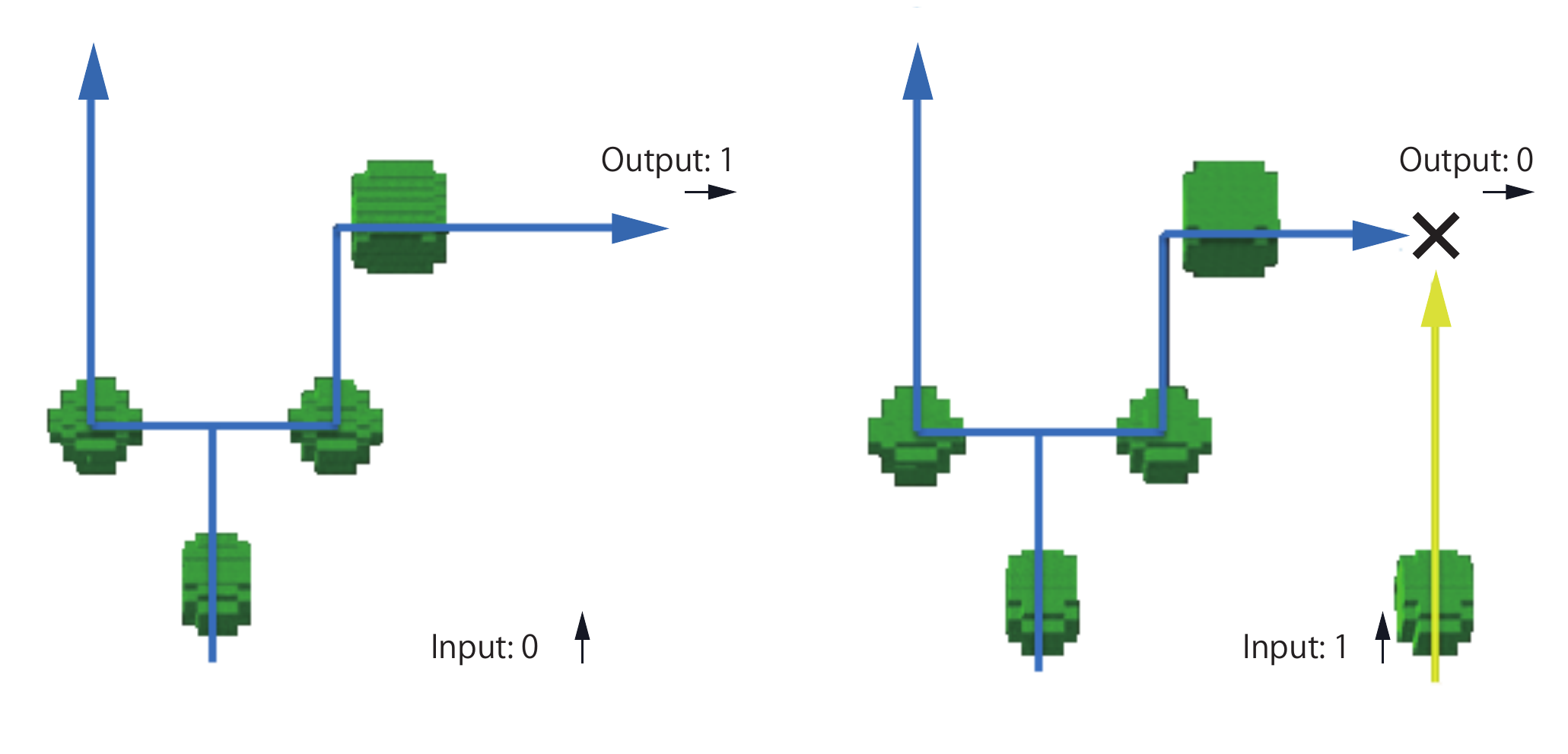}
\caption{NOT gate modules (input 0 and 1).}
\label{fig:not-pattern}
\end{center}
\end{figure}

Fig.~\ref{fig:not-pattern} is a NOT gate module. A duplicator is used as a kind of ``clocking" module and generate a bug. An input signal to the NOT gate is encoded by the existence of a bug. If the input is $0$ then the generated bug is used as the output. If the input is $1$ then the associated input bug B0 causes a collision to the generated bug. Almost all types of collisions of two bugs finally annihilate all live cells produced by the collision. Thus there is no output, i.e., $0$. 

\begin{figure}[h]
\begin{center}
\includegraphics[width=0.4\linewidth]{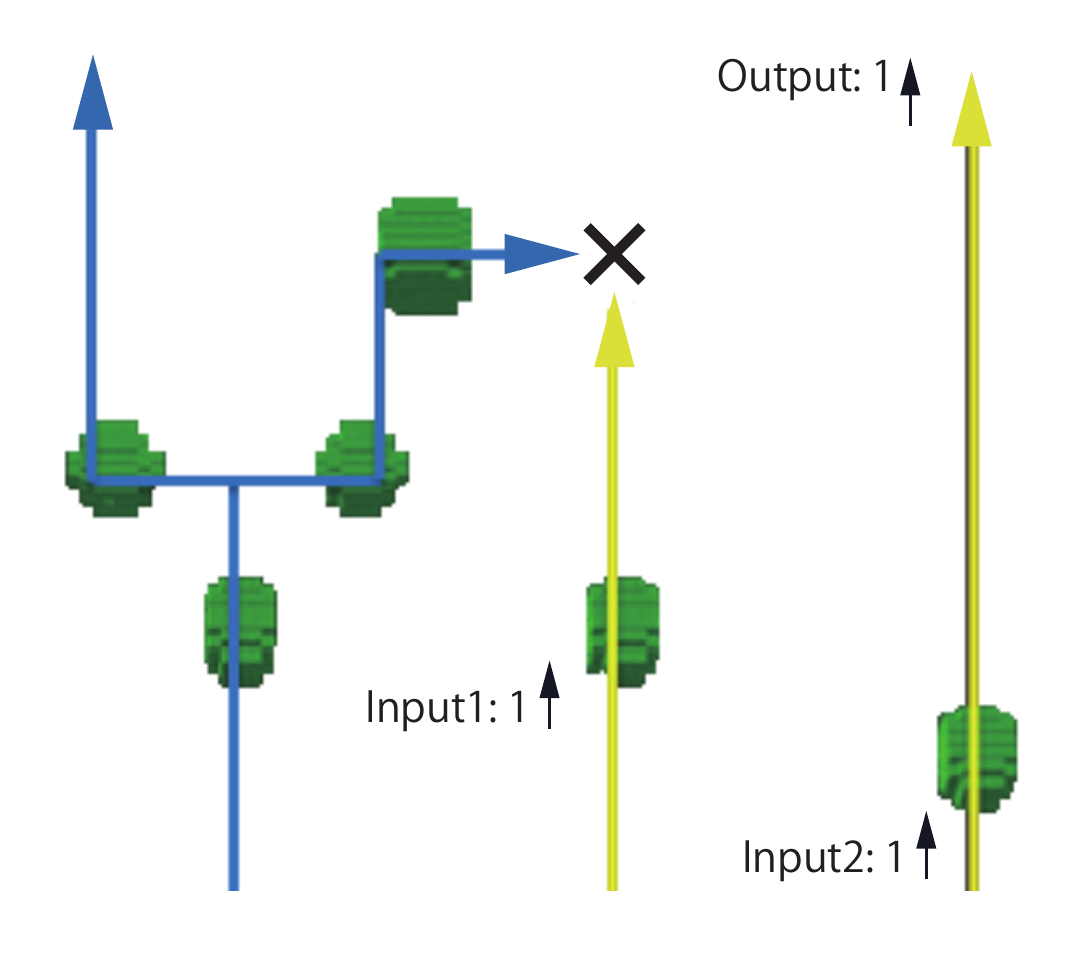}
\caption{An AND gate module (input 1 and 1).}
\label{fig:and-pattern}
\end{center}
\end{figure}

Fig.~\ref{fig:and-pattern} shows an AND gate module. It can be realized by the similar way. 
Only in the case that two bugs are placed as inputs 1 and 1, the input2 bug survives as its output. 

\subsection{Synchronizing issue}

The position of input and output bugs of a duplicator (a turning module) differs in z-coordinate. The value of the offset of a duplicator and turning module are $1$ and $6$ respectively. 
Each module flipped upside down is also effective, i.e., a flipped duplicator is also a duplicator, and a flipped (right) turning module can be used as a (left) turning module. 

\begin{figure}[h]
\begin{multicols}{2}
\begin{center}
\includegraphics[width=0.9\linewidth]{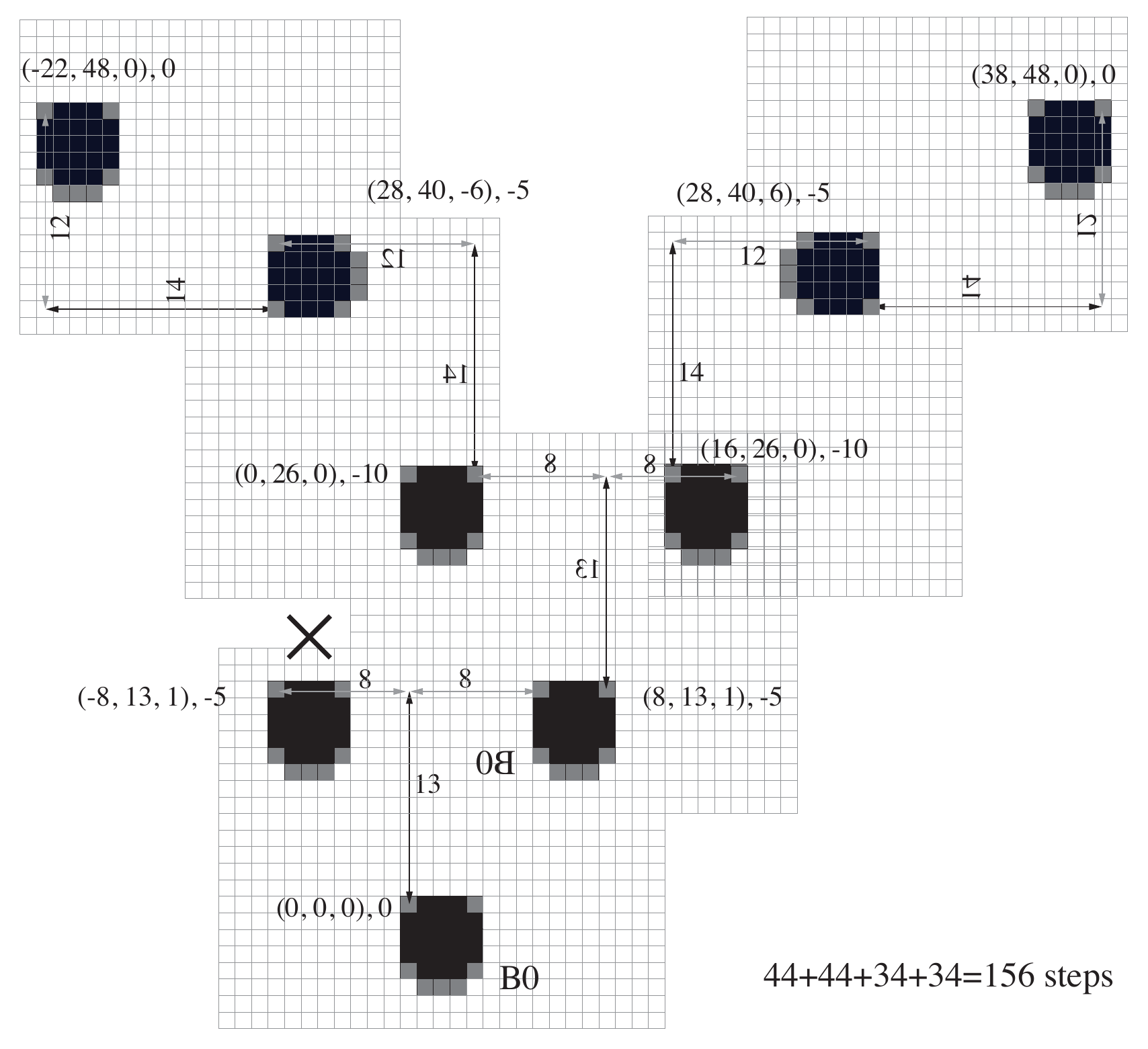}
\caption{A synchronized duplicator.}
\label{fig:sync-dup}
\end{center}

\begin{center}
\includegraphics[width=1.0\linewidth]{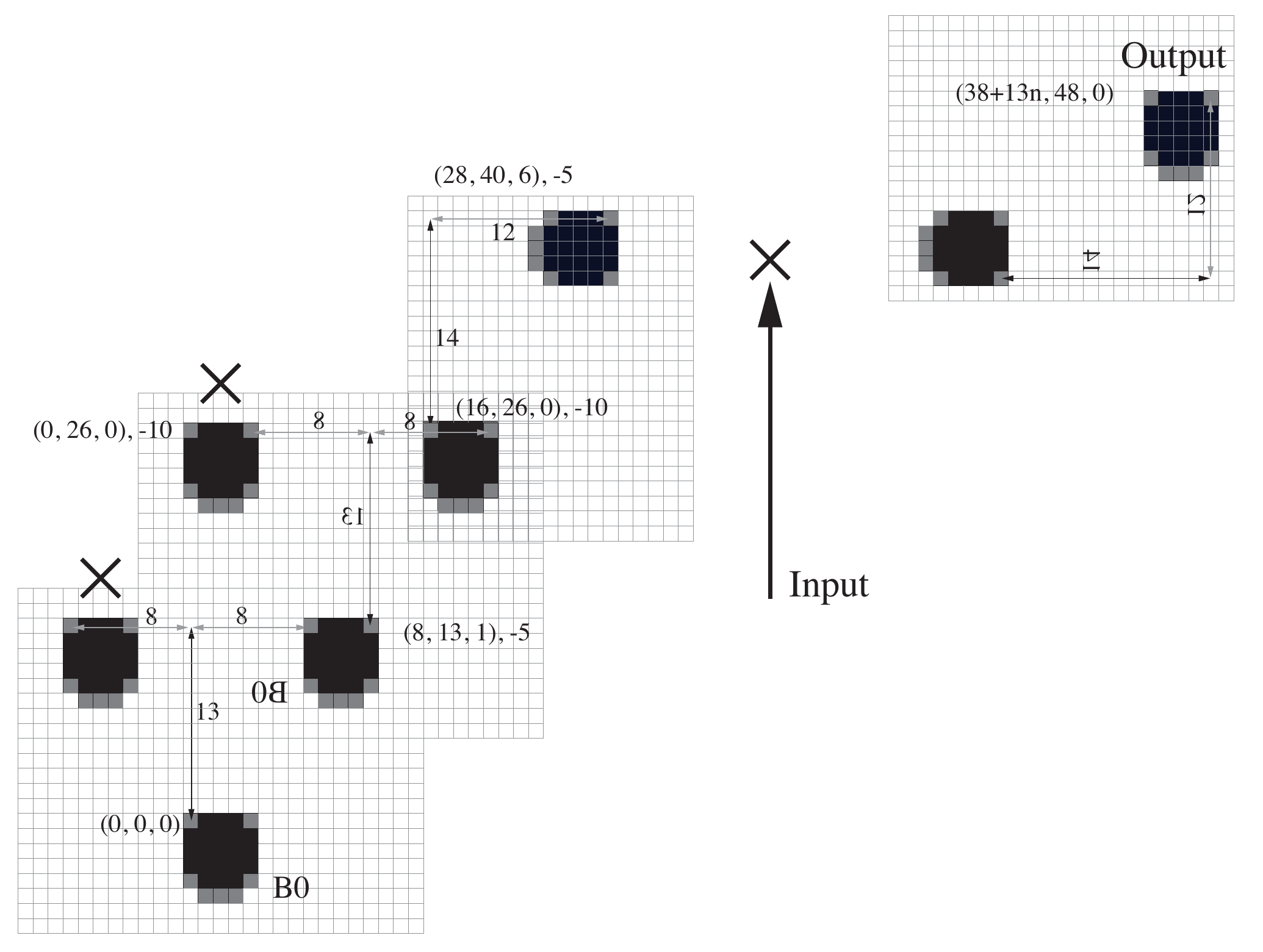}
\caption{A synchronized NOT gate.}
\label{fig:sync-not}
\end{center}

\end{multicols}
\end{figure}

The positions of duplicated bugs by the pattern in Fig.~\ref{fig:dupbug-schematic} is $(\pm 8, 13, 1)$ and the difference of their phase of period from the input bug is $-5$. Because another B0 pattern appear at 44-step later, but the original bug must be B0 pattern again in 39-step later. The change of the phase of period causes a serious problem for combining the pattern of gate modules. Because we need to change the phase of blinkers P and Q depending on the number of cascading modules. 
%
But the difference of the phase of period between input and output bugs of turning module pattern (Fig~\ref{fig:turn-schematic}) is ``fortunately" $5\  (=39-34)$. Thus we can reset the phase difference by combining the same number of both modules. 
Fig.~\ref{fig:sync-dup} illustrates the positions of input and output bugs of a phase synchronized duplicator. 

Next, we consider the logical gate patterns. 
In general, two inputs of an AND gate in Fig.~\ref{fig:and-pattern} should be strictly synchronized both in their positions and phases. But in our case, only the bug of input 2 is used as its output signal and the bug of input 1 is just used as an inhibitor of the bug generated by the duplicator. Because the annihilation of bugs by collisions occurs in almost all combination of phases, we do not need to synchronize two inputs. Thus an AND gate never changes the phase of input signal. We also do not need to control the position strictly. Of course, a bug with cross mark needs to be eliminated. But almost all types of collisions with a bug and a blinker P finally cause annihilation. But to avoid confliction of their transitional evolution, one need to leave enough space between each module.

A NOT gate in Fig.~\ref{fig:not-pattern} needs to be carefully considered because the output signal uses the bug generated from a clocking signal by a duplicator and the phase of the output bug is not synchronized. So we replace the previous NOT gate pattern with a synchronized version. Fig.~\ref{fig:sync-not} illustrates the input and output relation of bugs for an improved NOT gate. This layout also usable as a delay or changing the positions of bugs for wiring a circuit.

\section{Conclusion}

In this paper, we show that an LtL, $(4,102,133,102,142)$, is a candidate for weak universality. Because universal logic elements can be embedded and a necessary wiring method are shown, it should be possible to build a cell module for the rule 110 and placing it infinitely but periodically to simulate the evolution of the rule 110 cellular automaton. 
To the best of our knowledge, this rule is the first universal candidate among three-dimensional life-like and LtL rules. 

As in the case of two-dimensional LtL rule, Bosoco, a bug gun should be the most important module. It seems inevitable to achieve strong universality, i.e., universality with finite initial configuration. We strongly believe the existence of Bosco like rules in three-dimensional LtL. But we can't find any bug gun so far in three-dimension. 

\subsubsection*{Acknowledgments.} 

This work is supported by the JSPS KAKENHI Grant Numbers JP26330016, JP17K00015.

%
%

\end{document}